\documentclass[11pt,twoside]{article}
\usepackage{asp2010}

\resetcounters

\begin{document}

\title{Continuous upflow of material in an active region filament from the photosphere to the corona}
\author{Christoph Kuckein$^{1,2}$, Rebecca Centeno$^3$, and Valent\'in Mart\'inez Pillet$^1$
\affil{$^1$Instituto de Astrof\'isica de Canarias, C/ V\'ia L\'actea s/n, E-38205 La Laguna, Tenerife, Spain}
\affil{$^2$Departamento de Astrof\'isica, Universidad de La Laguna, E-38206 La Laguna, Tenerife, Spain}
\affil{$^3$High Altitude Observatory (NCAR), Boulder, CO 80301, USA}}

\begin{abstract}
Using spectropolarimetric data of an Active Region (AR) filament we have carried out inversions in order to infer vector magnetic fields in the photosphere (\ion{Si}{i} line) and in the chromosphere (\ion{He}{i} line). Our filament lies above the polarity inversion line (PIL) situated close to disk center and presents strong Zeeman-like signatures in both photospheric and chromospheric lines. Pore-like formations with both polarities are identified in the continuum under the PIL. The azimuth ambiguity is solved at both heights using the AZAM code. A comparison between the photospheric and chromospheric vector magnetic fields revealed that they are well aligned in some areas of the filament. However, especially at chromospheric heights, the magnetic field is mostly aligned with the dark threads of the filament. Velocity signatures indicating upflows of field lines are found at both heights. The combination of all these findings strongly suggests an emerging flux rope scenario.
\end{abstract}

\section{Introduction}
Filaments, also called prominences if observed outside the solar disk, consist of dense plasma that lays above polarity inversion lines (PILs). On disk observations using the \ion{He}{i} 10830 \AA\ multiplet or the H$\alpha$ spectral line allow us to identify filaments at chromospheric heights. We can distinguish between two types of filaments: active region (AR) and quiescent filaments. Their magnetic structure and formation are still not well understood and although new observations with better resolution are being carried out, the discussion between different models, e.g. twisted flux rope and sheared arcade models, still remain \citep[see][and references therein]{ck-mackay}.

In reviewing the literature, almost no studies of chromospheric velocity measurements in AR filaments have been made. On the other hand, quiescent filament threads have been measured recently by \citet{ck-chae} obtaining velocities of up to $\pm$15 km/s. In this paper we will present velocity flows as well as vector magnetic fields of an AR filament. 

\section{Observations and inversions}
We have analyzed spectropolarimetric data acquired with TIP (Tenerife Infrared Polarimeter) at the VTT (Tenerife, Spain) taken on July 3rd and 5th, 2005, of an AR filament (NOAA 10781) near disk center ($\mu = 0.97$ and $\mu = 0.92$ respectively). The observed spectral range spanned from 10825 to 10836 \AA, with a spectral sampling of \mbox{$\sim$ 11.1} m\AA/px, and include the chromospheric \ion{He}{i} 10830 \AA\ multiplet and photospheric \ion{Si}{i} 10827 \AA\ spectral line. Flat-field and dark current corrections as well as polarimetric calibration were carried out \citep{ck-collados99, ck-collados03}. Further details of the observations are described by \citet{ck-kuckein09}.

In order to improve the signal-to-noise ratio we applied a binning of 3px (along the scan and spectral axes) and 6px (along the slit) to the data. For the inversion of the full Stokes profiles of the \ion{He}{i} 10830 \AA\ multiplet we used a Milne-Eddington (ME) inversion code \citep[MELANIE;][]{ck-melanie}. This code computes the Zeeman-induced Stokes spectra in the incomplete Paschen-Back effect regime. \citet{ck-kuckein09} showed that the Stokes profiles are dominated by the Zeeman effect and that atomic polarization is almost absent in the filament. We therefore consider this ME inversion code as a good approximation to invert our data. 

The Stokes profiles of the photospheric \ion{Si}{i} 10827 \AA\ line were inverted with the SIR code \citep{ck-sir} under local thermodynamic equilibrium (LTE) conditions assuming an initial penumbral model atmosphere. 

\section{Vector magnetic field}

\begin{figure}[!t] 
 \plottwo{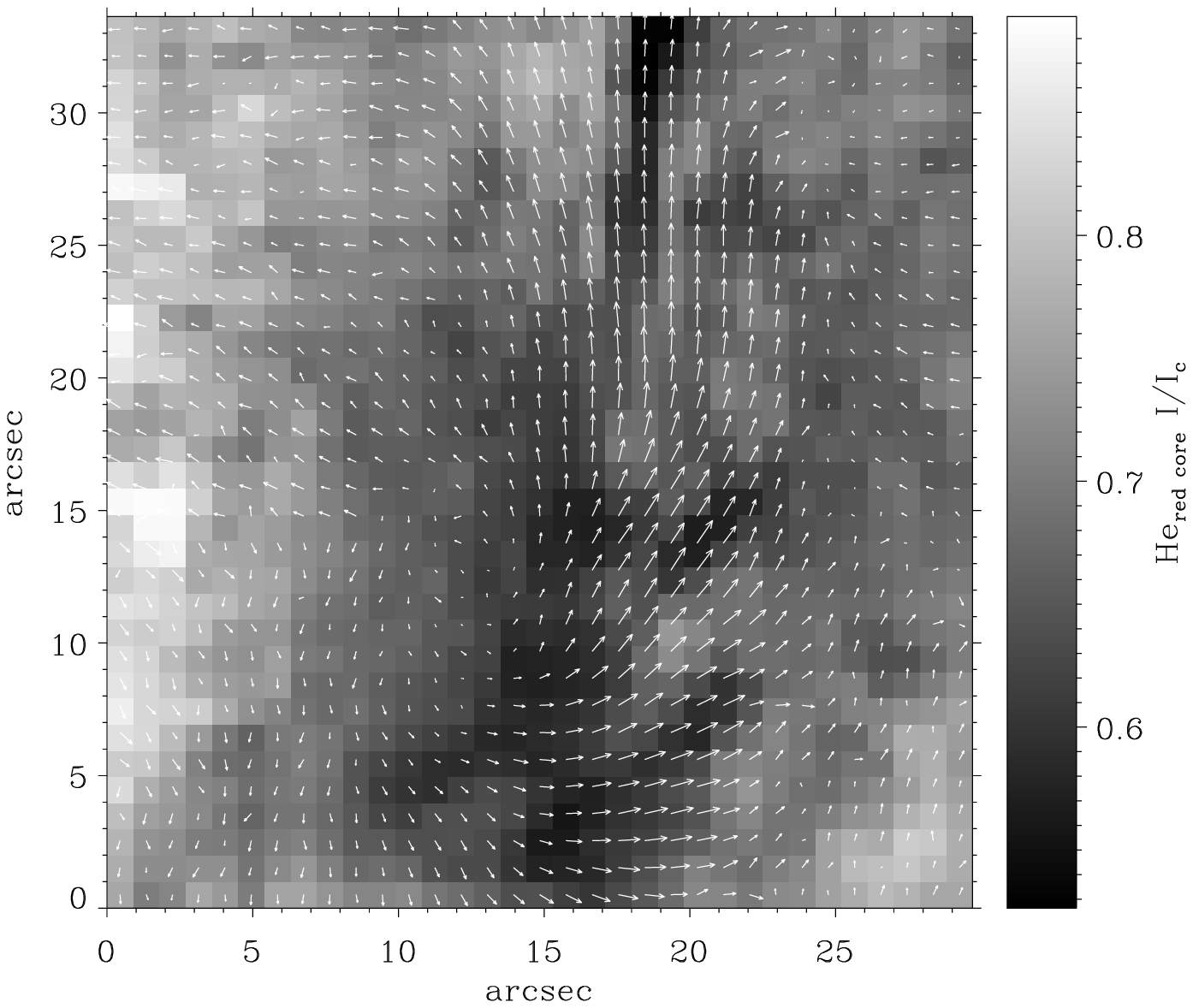}{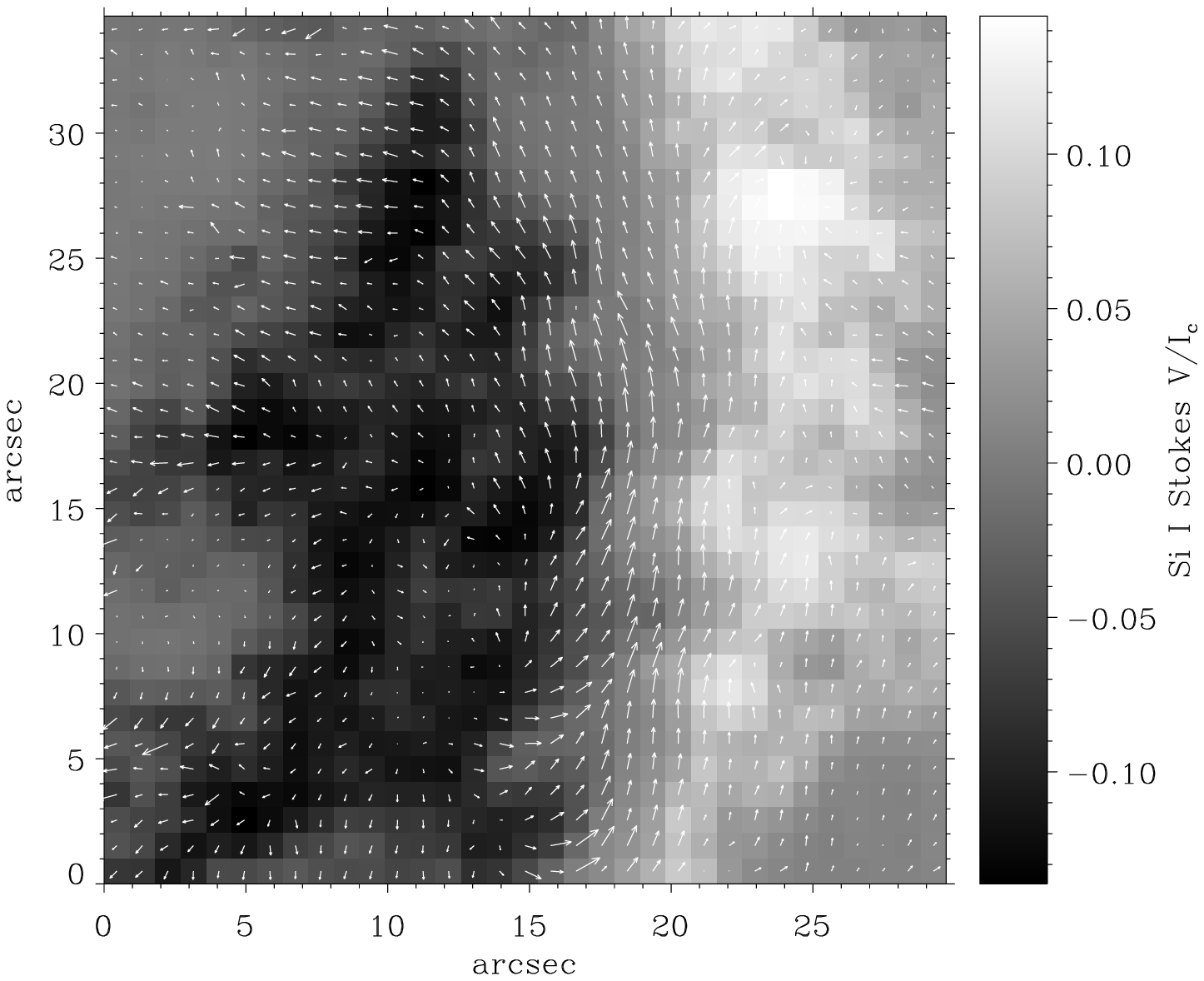}
 \caption{\textit{Left}: Chromospheric vector magnetic field over-plotted to the \ion{He}{i} red core continuum normalized intensity map. The arrows are well aligned with the dark threads. \textit{Right}: Photospheric magnetic field arrows over-plotted to the photospheric Stokes $V/I_\mathbf{c}$ image. A clear alignment with the PIL and a highly sheared magnetic field are identified.} 
\label{fig:ck-vmf}
\end{figure}

After the inversion we solved the 180$^{\circ}$ azimuth ambiguity using the AZAM code \citep{ck-azam} which was adapted to our TIP data. The vector magnetic fields of \ion{He}{i} and \ion{Si}{i}, at an optical depth of $\log \tau = -2$ for the latter, are presented in Figure \ref{fig:ck-vmf}. 

The chromospheric magnetic field is highly transversal, and forming an angle of almost $\sim$ 90$^\circ$ with respect to the filament axis in the lower part of the \textit{left} panel in Fig. \ref{fig:ck-vmf} while, in contrast, in the upper part it is parallel to the filament. It is likely therefore that we are seeing different heights in the layer of formation of \ion{He}{i}. Hence, the upper part would represent the axis of the filament while closer to the bottom we would be looking at dips of twisted magnetic field lines. Moreover, the chromospheric vector magnetic field is well aligned with the dark \ion{He}{i} threads of the filament and it is systematically less sheared than the photospheric one especially in the bottom part of the filament. The \textit{right} panel of Fig. \ref{fig:ck-vmf} presents the vector magnetic field inferred from the \ion{Si}{i} line which is well aligned with the PIL. At the top part of the filament, photospheric and chromospheric fields are almost parallel to each other. This behavior is common to the other data-sets of the same region taken on July 5th. Note that the distribution of polarities and the orientation of the \ion{He}{i} threads showed in Fig. \ref{fig:ck-vmf}, together with the azimuth of the field lines pointing upward, strongly suggest dips in a flux rope topology.

\section{Velocities}
\begin{figure}[!t] 
 \plottwo{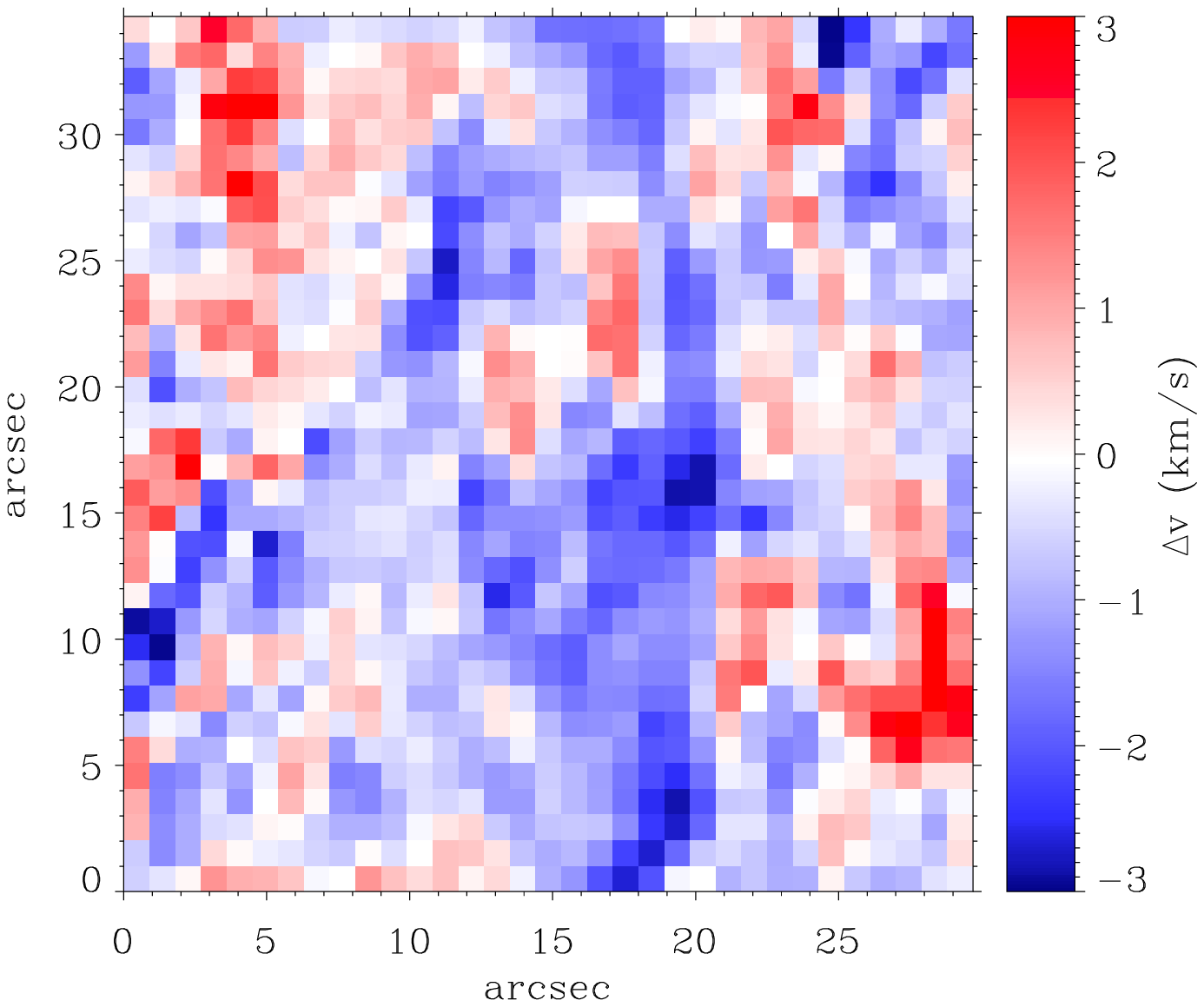}{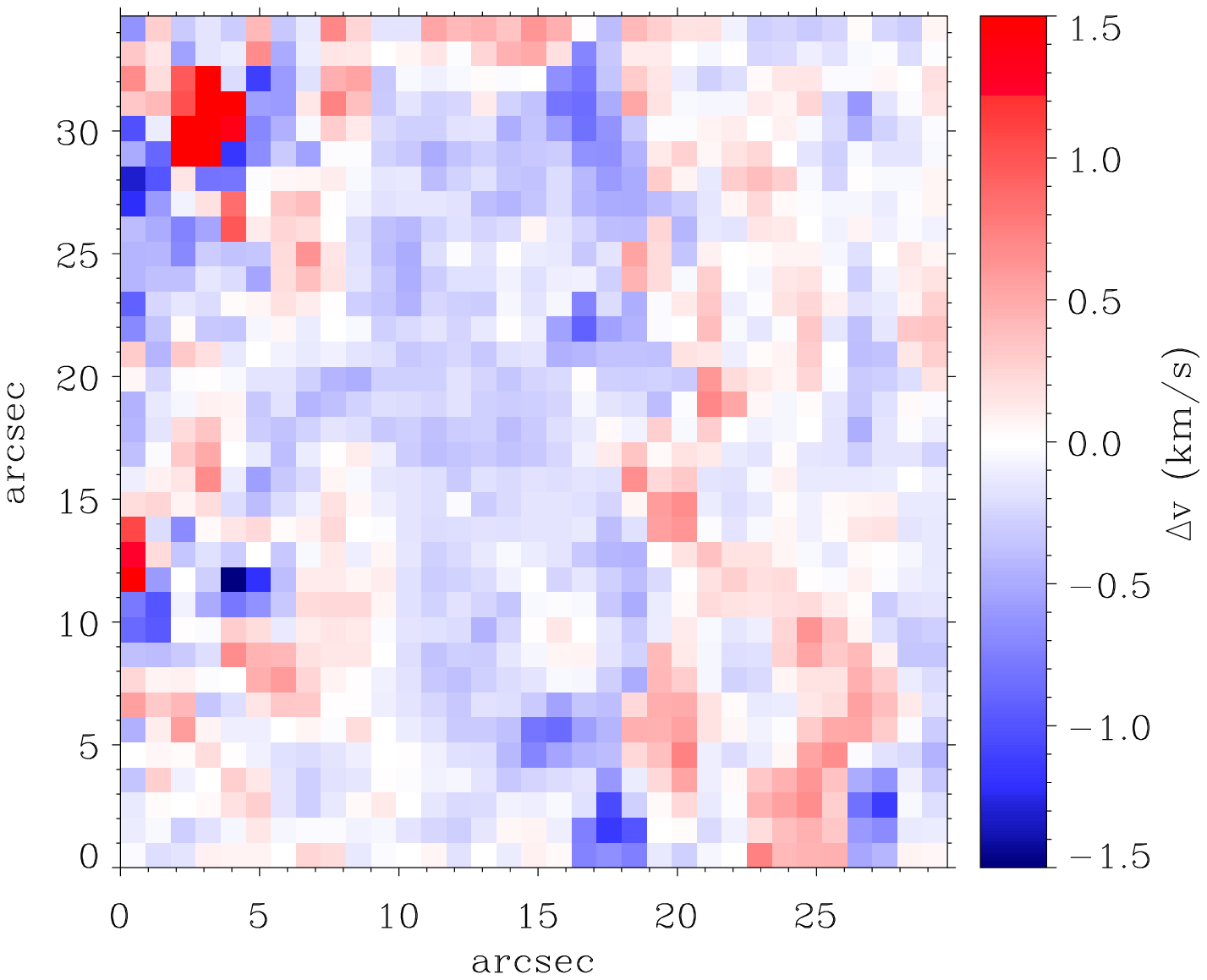}
 \caption{The \textit{left} (\textit{right}) panel shows the \ion{He}{i} (\ion{Si}{i}) velocity map, $\Delta v$, with respect to the mean faculae velocity of the map. Upflows (\textit{blue}) at both heights can be seen in the filament. Note that the color scale is not the same in both panels.} 
\label{fig:ck-vel}
\end{figure}

In order to infer line-of-sight (LOS) velocities as accurate as possible, we did a wavelength calibration of our data, correcting them for orbital motions which contribute to the wavelength shifts (Sun and Earth motions; gravity shift). To understand if the field lines in the filament are moving up or down, we refer the velocities in them to those in the faculae (which act as our zero of velocity). In Fig. \ref{fig:ck-vel} we present the LOS velocities ($\Delta v$) with respect to the faculae for the same field of view as Fig. \ref{fig:ck-vmf}. To make sure that we were measuring the velocities of the filament, we established the following criterion: $I_\mathrm{He}/I_\mathrm{c} < 0.6$, which corresponds to the dark threads of Fig. \ref{fig:ck-vmf}. At chromospheric heights, \textit{left} panel of Fig. \ref{fig:ck-vel}, the filament has a mean upward motion of $\sim 1.1$ km/s with peak velocities up to $2.8$ km/s. On the other hand, in the photosphere, \textit{right} panel, at an optical depth of $\log \tau = -2$ the velocities are much smaller but the plasma is still rising with a mean velocity of $0.1$ km/s. Interestingly, in all other maps of the 3rd and 5th of July we also find upflows. There is a continuous upflow of plasma in the filament that rises up to the corona. 

\section{Conclusions and further work}
In this paper we have presented vector magnetic fields as well as LOS velocities for two different heights of an AR filament. This study has found that generally the main axis of the filament is moving upward in the chromosphere as well as in the photosphere, being the latter velocity less significant.

The chromospheric vector magnetic field is well aligned with the \ion{He}{i} dark threads. Taking into account that the positive polarity is on the right side of the filament and the negative one on the left, a potential field solution would show field lines that point from right to left. But this is not the case as shown in Fig. \ref{fig:ck-vmf}. An implication of this inverse chromospheric vector magnetic field configuration is the presence of dips in a flux rope topology. 

It was also shown that the photospheric vector magnetic field is aligned with the PIL and has highly sheared magnetic field lines. 

All in all, an emerging flux rope scenario is strongly suggested from the combination of observed velocities and vector field distribution. Further work needs to be done to understand the formation and evolution of this AR filament. We will perform nonlinear force-free field (NLFFF) extrapolations using our inferred vector magnetic fields as boundary data.

\acknowledgements 
Based on observations made with the VTT operated on the island of Tenerife by the KIS in
the Spanish Observatorio del Teide of the Instituto de Astrof\'isica de Canarias. NCAR is sponsored by the National Science Foundation. Christoph Kuckein would like to thank the LOC of the SPW6 for the travel support.  

\bibliography{kuckein}

\end{document}